\documentclass[lettersize,journal]{IEEEtran} 
\usepackage{amsmath,amsfonts,amssymb}
\usepackage{algorithmic}
\usepackage[dvips]{graphicx}
\usepackage{verbatim}
\usepackage{caption}
\usepackage{subcaption}
\usepackage{cite}

\usepackage{lipsum}
\usepackage{changepage}
\usepackage{pdfpages}
\usepackage{color}

\usepackage{setspace}
\usepackage{bm}
\usepackage[ruled,vlined]{algorithm2e}

 \usepackage[labelformat=simple]{subcaption}

\IEEEoverridecommandlockouts

\begin{document}
%

\title{
    \LARGE{Wideband Coverage Enhancement for IRS-Aided Wireless Networks Based on Power Measurement}
    \vspace{-8pt}
}

%

\author{Ge Yan, 
        Lipeng Zhu,~\IEEEmembership{Member,~IEEE}, 
        He Sun,~\IEEEmembership{Member,~IEEE}, 
        Rui Zhang,~\IEEEmembership{Fellow,~IEEE}
        
        \vspace{-35pt}

\thanks{G. Yan is with the NUS Graduate School, National University of Singapore, Singapore 119077, and also with the Department of Electrical and Computer Engineering, National University of Singapore, Singapore 117583 (e-mail: geyan@u.nus.edu). 
L. Zhu, H. Sun, and R. Zhang are with the Department of Electrical and Computer Engineering, National University of Singapore, Singapore 117583 (email: zhulp@nus.edu.sg; sunele@nus.edu.sg; elezhang@nus.edu.sg). }
}


\maketitle

\IEEEpeerreviewmaketitle

\begin{abstract}
    By applying tunable phase shifts to incident waves via passive signal reflection, intelligent reflecting surface (IRS) can offer significant performance improvement for wireless communication systems. 
    To reap such performance gain, channel knowledge for IRS-cascaded links is generally required, which is practically challenging to acquire due to their high-dimensional and time-varying characteristics. 
    Conventional pilot-based channel estimation incurs excessive overhead due to the large number of reflecting elements, thus undermining the IRS efficiency, especially for wideband systems with frequency-selective fading channels. 
    To tackle this issue, we propose in this letter a power-measurement-based channel autocorrelation matrix estimation and coverage enhancement approach for IRS-aided orthogonal frequency division multiplexing (OFDM) systems. 
    Specifically, by estimating equivalent channel autocorrelation matrices of IRS-cascaded OFDM channels based on receive signal power and optimizing the IRS reflection vector based on them, the average coverage performance in the IRS-aided region is enhanced without the need for frequent reconfiguration of IRS reflection coefficients based on user instantaneous channels. 
    Simulation results validate the effectiveness of the proposed approach for improving the average channel gain over the coverage region. 
\end{abstract}

\vspace{-4pt}

\vspace{-8pt}
\section{INTRODUCTION}\label{sec:introduction}
    \IEEEPARstart{I}{ntelligent} reflecting surface (IRS) is deemed as a promising technology for future wireless networks due to its capability of reconfiguring wireless channels cost-efficiently~\cite{ref:PIEEE-IRS6G}. 
    However, to fully exploit the performance gains of IRS, channel state information (CSI) for IRS-cascaded channels is required, which poses great challenges in practice for CSI acquisition, especially for IRSs with a large number of reflecting elements and assisting wideband wireless systems with frequency-selective fading channels~\cite{ref:PIEEE-IRS6G}. 

    To tackle this challenge, extensive studies have been conducted to estimate the cascaded channel for IRS-aided systems. 
    In particular, conventional pilot-based channel estimation approaches are commonly employed, where additional pilot signals are sent from the base station (BS) to users with time-varying IRS reflection coefficients based on e.g., discrete Fourier transform (DFT) or Hadarmard matrix based codebooks~\cite{ref:ce-bf-discrete}. 
    Besides, the sparsity of channel paths in the angular domain was exploited in~\cite{ref:CE-IRS-double-sparsity} to apply compressed sensing techniques, which improves the efficiency of channel estimation. 
    However, in the current wireless communication standards~\cite{ref:3gpp:38.211}, pilot signals are dedicated only for channel estimation of the direct link between the BS and users. 
    Thus, significant modifications to communication protocols are required for implementing pilot-based channel estimation schemes in IRS-aided wireless systems. 
    To facilitate practical implementation of IRS, the authors in~\cite{ref:CSM, ref:acsm, ref:my-twc-ver-inst, ref:sun-he-wideband-twc} proposed to optimize IRS reflections based on received signal power measurements of users terminals, e.g., reference signal received power (RSRP)~\cite{ref:3gpp:36.214}, which can be readily obtained from existing wireless systems. 
    Specifically, the IRS reflection coefficients were designed by leveraging the statistical information embedded in received power measurements without the need for explicit channel estimation~\cite{ref:CSM, ref:acsm}. 
    In contrast, the autocorrelation matrix of the cascaded IRS channel was estimated in~\cite{ref:my-twc-ver-inst,ref:sun-he-wideband-twc} based on power measurements, which is then used for more efficient IRS reflection optimization. 
    
    In the aforementioned works, channel estimation and IRS reflection optimization are conducted based on instantaneous channels, which is challenging to implement under fast-varying channels. 
    Thus, a promising alternative approach is to design IRS reflection coefficients based on statistical CSI~\cite{ref:sun-he-coverage-twc} for enhancing the coverage performance over the IRS-aided region~\cite{ref:ma-bf-3d-cover}. 
    As such, the overall communication performance of users within the region over a long term can be improved with no need for real-time IRS configurations catering to instantaneous CSI. 
    To this end, an IRS-aided narrowband system was considered in~\cite{ref:sun-he-coverage-twc}, where the IRS-cascaded channels within the region are estimated by neural network based on received power measurements, which are utilized to design IRS reflection coefficients for maximizing the average receive signal-to-noise-ratio (SNR) in the region. 

    In this letter, we consider an IRS-aided orthogonal frequency division multiplexing (OFDM) system and study the wideband channel autocorrelation matrix estimation and coverage enhancement problem based on received power measurements. 
    Specifically, the channel autocorrelation matrices at several sampled locations within the region of interest are estimated given the power measurements, based on which the IRS reflection coefficients are designed to maximize the average receive SNR over the region. 
    By approximating the estimation problem into a least-square (LS) optimization problem with the rank-limited constraint, a wideband approximate-low-rank-approaching (W-ALRA) algorithm is proposed to estimate the channel autocorrelation matrix by progressively refining its rank. 
    Simulation results verify the effectiveness of the proposed W-ALRA algorithm, as well as its resulting coverage improvement compared to other benchmark schemes, for IRS-aided wideband communication systems.

\vspace{-6pt}
\section{System Model and Problem Formulation}\label{sec:system-model-and-problem-formulation}

    \vspace{-2pt}
    \subsection{System Model}\label{subsec:system-model}

\begin{figure}[t]
            \begin{center}
                \includegraphics[scale = 0.26]{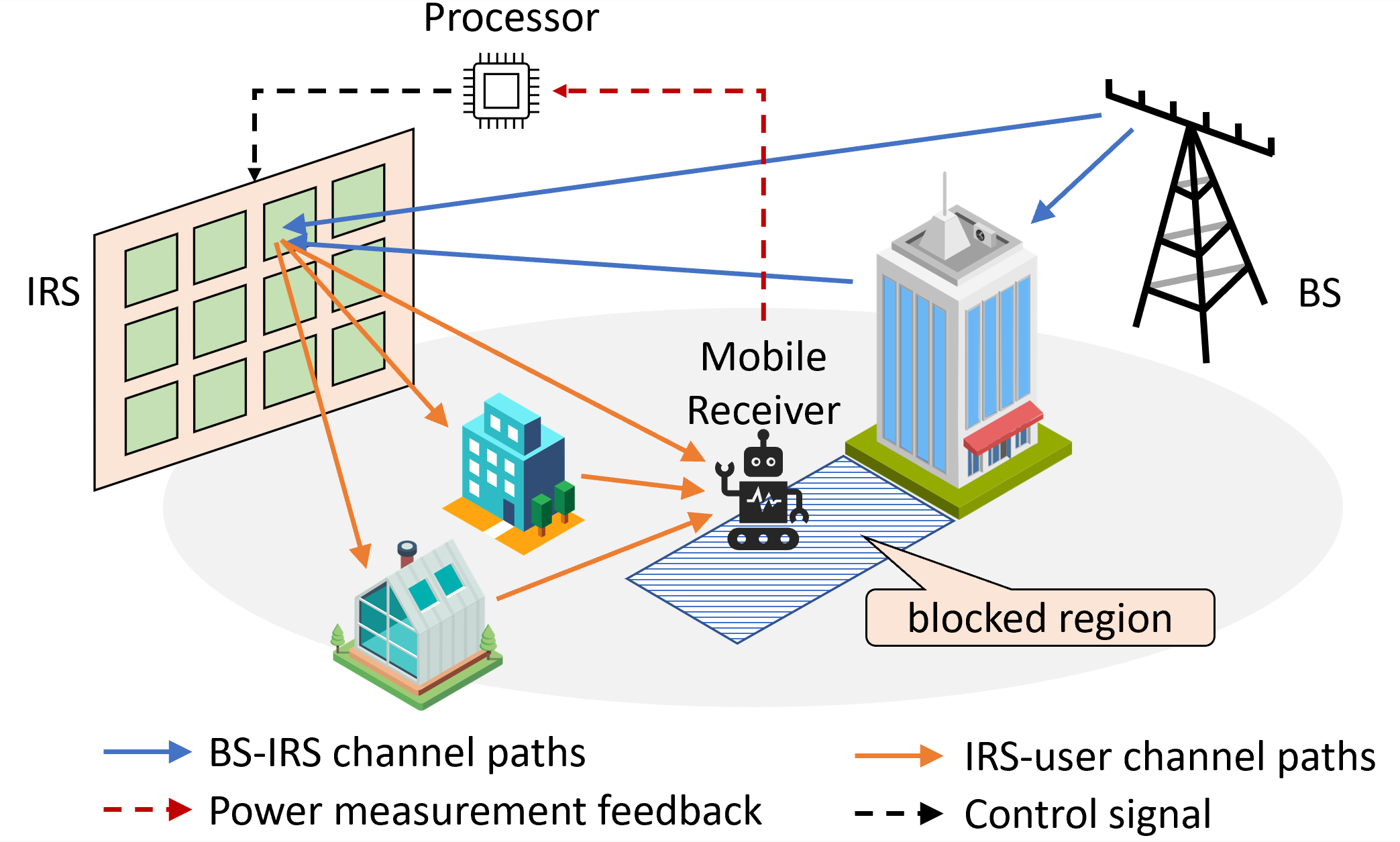}
                \vspace{-4pt}
                \caption{IRS-aided wideband coverage enhancement. }
                \label{fig:system-setup}
            \end{center}
            \vspace{-28pt}
        \end{figure}

        We consider a downlink OFDM system with a multi-antenna BS serving a particular region of interest. 
        Due to obstacles, the line-of-sight (LoS) paths from the BS to the region, denoted as $\mathcal{C}$, are severely blocked. 
        To improve the communication performance for users within $\mathcal{C}$, an IRS with ${N}$ reflecing elements is deployed to establish a reflected link between the BS and $\mathcal{C}$. 
        To this end, the BS needs to steer its beam towards the IRS so that more signal power can be reflected to $\mathcal{C}$ by the IRS. 
        Thus, the BS beamforming can be assumed to be fixed and the BS can be equivalently considered as having one single directional antenna. 
        To measure the coverage performance, a mobile receiver is deployed in the considered region for collecting the received signal power, as shown in Fig.~\ref{fig:system-setup}. 
        By denoting the phase shift introduced by the $n$-th reflecting element as ${v}_{n}$, $1\le n\le {N}$, we define the IRS reflection vector as $\boldsymbol{v} = [{v}_{1}, \ldots, {v}_{N}]^{T}\in\mathbb{C}^{N}$. 
        Besides, the number of bits controlling the phase shifts is denoted as $b$. 
        Thus, the reflection coefficient ${v}_{n}$ can only be configured as one of $2^{b}$ discrete values, i.e., ${v}_{n}\in\Phi_{b}\triangleq\{1, e^{j\Delta\theta}, \ldots, e^{j(2^{b} - 1)\Delta\theta}\}$, where $\Delta\theta = 2\pi/2^{b}$.

        Next, we denote $L_{g}$ as the maximum number of delay taps for the BS-IRS link and define the baseband equivalent channel from the BS to the $n$-th IRS reflecting element as $\boldsymbol{h}_{n}^{g} = [{h}_{n1}^{g}, \ldots, {h}_{nL_{g}}^{g}]^{T}\in\mathbb{C}^{L_{g}\times 1}$, $\forall n$. 
        Moreover, we define $L_{r}$ as the maximum number of delay taps for the IRS-receiver link within region $\mathcal{C}$.  
        Then, given any receiver in $\mathcal{C}$, the time-domain baseband equivalent channel from the $n$-th IRS reflecting element to it is denoted as $\boldsymbol{h}_{n}^{r} = [h_{n1}^{r}, \ldots, h_{nL_{r}}^{r}]^{T}\in\mathbb{C}^{L_{r}\times 1}$. 
        Accordingly, the time-domain cascaded channel for the BS-IRS-receiver link via the $n$-th IRS element is given by $\boldsymbol{h}_{n}^{c} = \boldsymbol{h}_{n}^{g}\circledast\boldsymbol{h}_{n}^{r}\in\mathbb{C}^{L\times 1}$, where $L = L_{g} + L_{r} - 1$ is the maximum number of delay taps and $\circledast$ represents convolution. 
        By further defining the equivalent channel matrix $\boldsymbol{H}_{c} = [\boldsymbol{h}_{1}^{c}, \ldots, \boldsymbol{h}_{N}^{c}]\in\mathbb{C}^{L\times{N}}$, the time-domain cascaded channel impulse response (CIR) of the BS-IRS-receiver link is written as 
        \begin{equation}\label{def:cascaded-cir}
            \boldsymbol{h} = \textstyle\sum_{n = 1}^{N}{
                \boldsymbol{h}^{c}_{n}\cdot{v}_{n}
            } = \boldsymbol{H}_{c}\cdot\boldsymbol{v}\in\mathbb{C}^{L\times 1}. 
        \end{equation}
        Define $M\gg L$ as the number of OFDM subcarriers and $\tilde{\boldsymbol{h}} = [\boldsymbol{h}^{T}, \boldsymbol{0}_{1\times (M - L)}]^{T}\in\mathbb{C}^{M\times 1}$ as the zero-padded CIR vector, where $\boldsymbol{0}_{a\times b}$ is a zero matrix of dimension $a\times b$. 
        Thus, the cascaded channel frequency response (CFR) of the BS-IRS-receiver link, denoted by $\boldsymbol{h}_{f} = [h_{f, 1}, \ldots, h_{f, M}]^{T}\in\mathbb{C}^{M\times 1}$, is given by 
        \begin{equation}\label{def:cfr-vector}
            \boldsymbol{h}_{f} = \boldsymbol{F}\tilde{\boldsymbol{h}} = \boldsymbol{F}\left[
                \begin{array}{c}
                    \boldsymbol{H}_{c} \\
                    \boldsymbol{0}_{(M - L)\times {N}}
                \end{array}
            \right]\boldsymbol{v}, 
        \end{equation}
        where $\boldsymbol{F}$ is the $M\times M$ DFT matrix. 
        Without loss of generality, we assume that each subcarrier is allocated with equal transmit power ${P}_{0}$ for convenience. 
        As such, the frequency-domain received signal $\boldsymbol{y}\in\mathbb{C}^{M\times 1}$ at the receiver can be expressed as 
        \begin{equation}\label{def:receive-signal-freq-domain}
            \boldsymbol{y} = \mathrm{Diag}(\boldsymbol{s})\boldsymbol{h}_{f} + \boldsymbol{z}, 
        \end{equation}
        where $\mathrm{Diag}(\boldsymbol{s})$ is the diagnol matrix of an OFDM symbol $\boldsymbol{s} = [s_{1}, \ldots, s_{M}]^{T}\in\mathbb{C}^{M\times 1}$ with $|s_{m}|^{2} = {P}_{0}$, $\forall m$, and $\boldsymbol{z} = [{z}_{1}, \ldots, {z}_{M}]^{T}\sim\mathcal{CN}(\boldsymbol{0}_{M\times 1}, \sigma^{2}\boldsymbol{I}_{M})\in\mathbb{C}^{M\times 1}$ is the receiver noise vector with average power $\sigma^{2}$ for each subcarrier. 
        Thus, the receive signal power of one OFDM symbol is given by 
        \begin{equation}\label{def:receive-signal-power}
            p_{r} = \|\boldsymbol{y}\|_{2}^{2} = \textstyle\sum_{m = 1}^{M}{
                \left|h_{f, m}{s}_{m} + z_{m}\right|^{2}
            } = {P}_{0}\|\boldsymbol{h}_{f}\|_{2}^{2} + Z, 
        \end{equation}
        where $Z = \sum_{m = 1}^{M}{|z_{m}|^{2} + 2\mathrm{Re}(h_{f, m}s_{m}\cdot z_{m}^{*})}$ is a random variable with $\mathbb{E}[Z] = M\sigma^{2}$. 
        By defining $\boldsymbol{V} = \boldsymbol{v}\boldsymbol{v}^{H}\in\mathbb{C}^{{N}\times{N}}$, we have $\|\boldsymbol{h}_{f}\|_{2}^{2} = \boldsymbol{v}^{H}\boldsymbol{R}_{c}\boldsymbol{v} = \mathrm{tr}(\boldsymbol{R}_{c}\boldsymbol{V})$, where $\boldsymbol{R}_{c}\in\mathbb{C}^{{N}\times{N}}$ is the wideband channel autocorrelation matrix defined as 
        \begin{equation}
            \boldsymbol{R}_{c} \triangleq \left[
                \begin{array}{c}
                    \boldsymbol{H}_{c} \\
                    \boldsymbol{0}_{(M - L)\times {N}}
                \end{array}
            \right]^{H}\boldsymbol{F}^{H}\boldsymbol{F}\left[
                \begin{array}{c}
                    \boldsymbol{H}_{c} \\
                    \boldsymbol{0}_{(M - L)\times {N}}
                \end{array}
            \right] = \boldsymbol{H}_{c}^{H}\boldsymbol{H}_{c}, 
        \end{equation}
        with $\mathrm{rank}(\boldsymbol{R}_{c})\le{L}\le{M}$. 
        Then, $p_{r}$ can be equivalently rewritten as 
        \begin{equation}\label{eqn:receive-signal-power-simplified}
            p_{r} = {P}_{0}\mathrm{tr}(\boldsymbol{R}_{c}\boldsymbol{V}) + Z. 
        \end{equation}

    \vspace{-16pt}
    \subsection{Coverage Enhancement}\label{subsec:coverage-enhancement}
        To enhance the coverage over $\mathcal{C}$, IRS reflection coefficients are designed to improve the average receive SNR within $\mathcal{C}$, or equivalently, the average channel power gain, i.e., $\bar{\gamma} \triangleq \boldsymbol{v}^{H}\mathbb{E}[\boldsymbol{R}_{c}]\boldsymbol{v}$. 
        Specifically, the mobile receiver moves to $K$ sampled locations within $\mathcal{C}$ for measuring the received signal power, each with the corresponding wideband channel autocorrelation matrix denoted as $\boldsymbol{R}_{c, k}$, $1\le k\le K$. 
        By estimating $\boldsymbol{R}_{c, k}$, $\forall k$ as $\hat{\boldsymbol{R}}_{c, k}$ based on received power measurements, $\bar{\gamma}$ can be approximated by the average channel power gain over the $K$ locations~\cite{ref:sun-he-coverage-twc}, denoted as $\hat{\gamma}$, i.e., $\bar{\gamma} \approx \hat{\gamma}\triangleq\boldsymbol{v}^H\hat{\boldsymbol{R}}_{c}\boldsymbol{v}$, where $\hat{\boldsymbol{R}}_{c} \triangleq \frac{1}{K}\sum_{k = 1}^{K}{\hat{\boldsymbol{R}}_{c, k}}$. 
        Then, the IRS reflection vector $\boldsymbol{v}$ is optimized to improve the coverage performance as follows:
        \begin{equation}\label{prob:average-snr-boost-multiary}
            \max_{\boldsymbol{v}} ~\boldsymbol{v}^{H}\hat{\boldsymbol{R}}_{c}\boldsymbol{v}, ~~\mathrm{s.t.} ~v_{n}\in\Phi_{b}, ~\forall n. 
        \end{equation}
        Given knowledge of $\hat{\boldsymbol{R}}_{c}$, problem~\eqref{prob:average-snr-boost-multiary} can be easily solved by applying semidefinite relaxation (SDR)~\cite{ref:sun-he-coverage-twc}. 
        However, the challenge for the considered coverage enhancement lies in the unkown $\hat{\boldsymbol{R}}_{c}$, which motivates our proposed wideband channel autocorrelation matrix estimation. 
        In the next subsection, power measurements at $K$ sampled locations are defined, based on which the wideband channel autocorrelation estimation problem is formulated.

    \vspace{-10pt}
    \subsection{Power Measurement and Problem Formulation}\label{subsec:power-measurements-and-prob-form}
        To estimate the wideband channel autocorrelation matrices $\boldsymbol{R}_{c, k}$, $\forall k$, the receive signal power values at each location are recorded with IRS randomly varying its reflection coefficients $\boldsymbol{v}$ for $T_{p}$ times, yielding $T_{p}$ power measurements at each location. 
        Each measurement is obtained by averaging over $J$ receive signal power samples over time with $\boldsymbol{v}$ fixed. 
        Specifically, for the $t$-th measurement, denote $\boldsymbol{v}_{t}$ as the IRS reflection vector, while $q_{k, t}$ is the power measurement obtained at the $k$-th location, $1\le t\le T_{p}$. 
        According to~\eqref{def:receive-signal-power} and~\eqref{eqn:receive-signal-power-simplified} with $\boldsymbol{V}_{t} = \boldsymbol{v}_{t}\boldsymbol{v}_{t}^{H}\in\mathbb{C}^{{N}\times{N}}$, $\forall t$, we have 
        \begin{subequations}\label{def:power-measurement}
            \begin{align}
                q_{k, t} & = {J}^{-1}\textstyle\sum_{i = 1}^{J}{
                    \textstyle\sum_{m = 1}^{M}{
                        \big|{h}_{f, km}^{t}\cdot{s}_{m} + {z}_{km, i}^{t}\big|^{2}
                    }
                } \\
                & = {P}_{0}\mathrm{tr}(\boldsymbol{R}_{c, k}\boldsymbol{V}_{t}) + J^{-1}\textstyle\sum_{i = 1}^{J}{
                    Z_{ki}^{t}
                }, 
            \end{align}
        \end{subequations}
        where ${h}_{f, km}^{t}$ is the CFR coefficient of the $m$-th subcarrier for the $t$-th power measurement at the $k$-th location, ${z}_{km, i}^{t}$ is the corresponding noise for the $i$-th received signal power value, $\forall i$, and $Z_{ki}^{t} = \sum_{m = 1}^{M}{|z_{km, i}^{t}|^{2} + 2\mathrm{Re}[{h}_{f, km}^{t}{s}_{m}({z}_{km, i}^{t})^{*}]}$, $\forall i$, are independent and identically distributed (i.i.d.) random noise power with $\mathbb{E}[Z_{ki}^{t}] = M\sigma^{2}$. 
        By defining $\bar{Z}_{k, t} = \frac{1}{J}\sum_{i = 1}^{J}{Z_{ki}^{t}}$ as the average noise power, we have $q_{k, t} = {P}_{0}\mathrm{tr}(\boldsymbol{R}_{c, k}\boldsymbol{V}_{t}) + \bar{Z}_{k, t}$. 
        For sufficiently large $J$, $\bar{Z}_{k, t}$ can be approximated as a complex Gaussian random variable, according to the central limit theorem, with $\mathbb{E}[\bar{Z}_{k, t}] = M\sigma^{2}$. 
        Thus, $q_{k, t}$ can be approximately regarded as a realization from a complex Gaussian distribution with mean ${P}_{0}\mathrm{tr}(\boldsymbol{R}_{c, k}\boldsymbol{V}_{t}) + M\sigma^{2}$. 

        Therefore, matrix $\boldsymbol{R}_{c, k}$ can be estimated by finding a low-rank matrix $\boldsymbol{R}$ to minimize the errors between the measured power $q_{k, t}$ and the expectated power, i.e., ${P}_{0}\mathrm{tr}(\boldsymbol{R}\boldsymbol{V}_{t}) + M\sigma^{2}$. 
        However, the IRS reflection vector $\boldsymbol{v}$ is a real vector for $b = 1$, while it is a complex vector for $b\ge 2$, which leads to different estimation mechanisms\footnote{For $b = 1$, the imaginary part of $\boldsymbol{R}$ cannot be distinguished based on power measurements. Thus, only the real part of $\boldsymbol{R}$ is estimated for $b = 1$, which is sufficient for coverage boost in this case, while the whole matrix $\boldsymbol{R}$ can be estimated for $b\ge 2$. More details can be found in Section III of~\cite{ref:my-twc-ver-inst}. }. 
        Thus, these two cases are handled separately in the following. 
        Specifically, for $b\ge 2$, the complex matrix $\boldsymbol{R}_{c, k}$ can be solved via the following optimization problem by assuming its rank as $D$, 
        \begin{subequations}\label{prob:channel-autocorr-est-multiary}
            \begin{align}
                \mathcal{P}_{b}^{k}(D): ~~ & \min_{\boldsymbol{R}} ~ \textstyle\sum_{t = 1}^{T_{p}}{
                    \left|
                        {P}_{0}\mathrm{tr}(\boldsymbol{R}\boldsymbol{V}_{t}) + M\sigma^{2} - q_{k, t}
                    \right|^{2}
                } \tag{\ref{prob:channel-autocorr-est-multiary}} \\
                & ~\mathrm{s.t.} ~ \boldsymbol{R}\in\mathbb{S}^{N}_{+}, ~\mathrm{rank}(\boldsymbol{R})\le{D}, \label{prob:channel-autocorr-est-multiary-constraints}
            \end{align}
        \end{subequations}
        where $\mathbb{S}_{+}^{N}$ is the set of $N\times N$ complex semidefinite matrices. 
        Although the exact value of $D_{k}\triangleq\mathrm{rank}(\boldsymbol{R}_{c, k})$ is usually unknown a priori, it can be relaxed to $M$ or progressively refined as a hyperparameter, which will be specified later. 
        Given the estimated results $\hat{\boldsymbol{R}}_{c, k}$, $\forall k$, the coverage performance can be maximized by optimizing $\boldsymbol{v}$ via 
        problem~\eqref{prob:average-snr-boost-multiary}. 
        On the other hand, for $b = 1$, only the real part of $\boldsymbol{R}_{c, k}$ is estimated, denoted as $\boldsymbol{R}_{r, k} = \mathrm{Re}(\boldsymbol{R}_{c, k})$, which is symmetric semidefinite and satisfies $\mathrm{rank}(\boldsymbol{R}_{r, k})\le 2D_{k}$. 
        Moreover, it was verified in~\cite{ref:my-twc-ver-inst} that, given any Hermitian matrix $\boldsymbol{A}$, we have $\boldsymbol{v}^{H}{\boldsymbol{A}}\boldsymbol{v} = \boldsymbol{v}^{H}\mathrm{Re}(\boldsymbol{A})\boldsymbol{v}$ for arbitrary IRS reflection vector $\boldsymbol{v}$ with $b = 1$. 
        Therefore, we have $q_{k, t} = {P}_{0}\mathrm{tr}(\boldsymbol{R}_{c, k}\boldsymbol{V}_{t}) + \bar{Z}_{k, t} = {P}_{0}\mathrm{tr}(\boldsymbol{R}_{r, k}\boldsymbol{V}_{t}) + \bar{Z}_{k, t}$ and $\boldsymbol{R}_{r, k}$ can be estimated via the following problem, 
        \begin{subequations}\label{prob:channel-autocorr-est-binary}
            \begin{align}
                \mathcal{P}_{1}^{k}(D): ~~ & \min_{\boldsymbol{R}_{r}} ~ \textstyle\sum_{t = 1}^{T_{p}}{
                    \left|
                        {P}_{0}\mathrm{tr}(\boldsymbol{R}_{r}\boldsymbol{V}_{t}) + M\sigma^{2} - q_{k, t}
                    \right|^{2}
                } \tag{\ref{prob:channel-autocorr-est-binary}} \\
                & ~\mathrm{s.t.} ~ \boldsymbol{R}_{r}\in\mathbb{M}^{N}_{+}, ~\mathrm{rank}(\boldsymbol{R}_{r})\le{2D}, \label{prob:channel-autocorr-est-binary-constraints}
            \end{align}
        \end{subequations}
        where $\mathbb{M}_{+}^{N}$ is the set of $N\times N$ real symmetric semidefinite matrices. 
        After obtaining the estimated result as $\hat{\boldsymbol{R}}_{r, k}$, $\forall k$, we have $\hat{\gamma} = \boldsymbol{v}^{H}\hat{\boldsymbol{R}}_{r}\boldsymbol{v}$, where $\hat{\boldsymbol{R}}_{r}\triangleq\frac{1}{K}\sum_{k = 1}^{K}{\hat{\boldsymbol{R}}_{r, k}}$. 
        Therefore, vector $\boldsymbol{v}$ can be optimized via solving problem~\eqref{prob:average-snr-boost-multiary} by replacing $\hat{\boldsymbol{R}}_{c}$ with $\hat{\boldsymbol{R}}_{r}$. 
        In the following, the proposed W-ALRA algorithms to solve problems~\eqref{prob:channel-autocorr-est-multiary} and~\eqref{prob:channel-autocorr-est-binary} are presented, along with the progressive refinement of $D$.

\vspace{-8pt}
\section{Proposed Solutions}\label{sec:Proposed}
    In this section, the W-ALRA algorithm is first illustrated to solve problem~\eqref{prob:channel-autocorr-est-multiary} for $b\ge 2$ and problem~\eqref{prob:channel-autocorr-est-binary} for $b = 1$ given $D$, respectively, after which the progressive refinement of $D$ is introduced. 

    \vspace{-8pt}
    \subsection{Case of $b\ge 2$ given $D$}
    \label{subsec:case-multiary}
        For $b\ge 2$, matrix $\boldsymbol{R}_{c, k}$ is estimated via problem $\mathcal{P}_{b}^{k}(D)$ given in~\eqref{prob:channel-autocorr-est-multiary}. 
        To handle the non-convex constraint $\mathrm{rank}(\boldsymbol{R})\le D$, it is incorporated into a penalty term and added to the objective function. 
        Specifically, we minimize the squared Euclidean distance between $\boldsymbol{R}$ and a rank-$M$ matrix $\boldsymbol{G}\in\mathbb{C}^{N\times N}$, denoted by $d_{e}^{2}(\boldsymbol{R}, \boldsymbol{G}) = \|\boldsymbol{R} - \boldsymbol{G}\|_{F}^{2}$, while accounting for the power measurement errors. 
        Specifically, $\mathcal{P}_{b}^{k}(D)$ is relaxed to the following problem:
        \begin{subequations}\label{prob:cov-est-dist-min-robust-multiary}
            \begin{align}
                & \mathop{\min_{\boldsymbol{R}, \boldsymbol{G}}}\ d_{e}^{2}(\boldsymbol{R}, \boldsymbol{G}) + \rho\textstyle\sum_{t = 1}^{T_{p}}{
                    \left|
                        {P}_{0}\mathrm{tr}(\boldsymbol{R}\boldsymbol{V}_{t}) + M\sigma^{2} - q_{k, t}
                    \right|^2
                } \tag{\ref{prob:cov-est-dist-min-robust-multiary}} \\
                & ~ \mathrm{s.t.} \ \boldsymbol{R}\in\mathbb{S}_{+}^{N}, ~\mathrm{rank}(\boldsymbol{G})\le{D}, \label{prob:cov-est-dist-min-robust-multiary-hermitian}
            \end{align}
        \end{subequations}
        where $\rho > 0$ is the penalty parameter. 
        By applying eigenvalue decomposition, $\boldsymbol{G}$ can be equivalently written as
        \begin{equation}
            \boldsymbol{G} = \textstyle\sum_{l = 1}^{D}\mu_{l}\boldsymbol{x}_{l}\boldsymbol{x}_{l}^{H} = \boldsymbol{X}\mathrm{Diag}(\boldsymbol{\mu})\boldsymbol{X}^{H}, 
        \end{equation}
        where $\boldsymbol{x}_{l}\in\mathbb{C}^{N\times 1}$, $\forall l$, are the normalized eigenvectors of $\boldsymbol{G}$, $\mu_{l}$, $\forall l$, are the corresponding eigenvalues, $\boldsymbol{X} = [\boldsymbol{x}_{1}, \ldots, \boldsymbol{x}_{D}]\in\mathbb{C}^{N\times D}$, and $\boldsymbol{\mu} = [\mu_{1}, \ldots, \mu_{D}]^{T}\in\mathbb{R}^{D\times 1}$. 
        Furthermore, we relax constraint $\boldsymbol{R}\in\mathbb{S}_{+}^{N}$ to $\boldsymbol{R}^{H} = \boldsymbol{R}$ and problem~\eqref{prob:cov-est-dist-min-robust-multiary} can be approximated by 
        \begin{subequations}\label{prob:cov-est-dist-min-multiary-relaxed}
            \allowdisplaybreaks
            \begin{align}
                \begin{split}
                    & \begin{aligned}[t]
                        \mathop{\min_{\boldsymbol{R}, \boldsymbol{\mu}, \boldsymbol{X}}}\ & \varphi_{k}\left(
                            \boldsymbol{R}, \boldsymbol{\mu}, \boldsymbol{X}
                        \right) = \left\|
                            \boldsymbol{R} - \boldsymbol{X}\mathrm{Diag}(\boldsymbol{\mu})\boldsymbol{X}^{H}
                        \right\|_F^2 \\
                        & ~~ + \rho\textstyle\sum_{t = 1}^{T_{p}}{
                            \left|
                                {P}_{0}\mathrm{tr}(\boldsymbol{R}\boldsymbol{V}_{t}) + M\sigma^{2} - q_{k, t}
                            \right|^2
                        }
                    \end{aligned} 
                \end{split} 
                \tag{\ref{prob:cov-est-dist-min-multiary-relaxed}} \\
                \begin{split}
                    & ~ \mathrm{s.t.} \ \boldsymbol{R}^H = \boldsymbol{R}, ~\boldsymbol{X}^{H}\boldsymbol{X} = \boldsymbol{I}_{D}, \label{prob:cov-est-dist-min-multiary-relaxed-hermitian}
                \end{split}
            \end{align}
        \end{subequations}
        where the objective function $\varphi_{k}(\boldsymbol{R}, \boldsymbol{\mu}, \boldsymbol{X})$ can be viewed as a penalized distance function between $\boldsymbol{R}$ and $\boldsymbol{G}$ given $q_{k, t}$, $\forall t$. 
        To solve problem~\eqref{prob:cov-est-dist-min-multiary-relaxed}, $\boldsymbol{R}$, $\boldsymbol{\mu}$, and $\boldsymbol{X}$ are optimized alternately. 
        Given $\boldsymbol{R}$, it can be easily verified that the optimal $\mu_{l}$ and $\boldsymbol{x}_{l}$ can be solved as the $l$-th largest eigenvalue of $\boldsymbol{R}$ and the corresonding normalized eigenvector of $\boldsymbol{R}$, respectively, $\forall l$. 
        Given $\boldsymbol{X}$, the optimal $\boldsymbol{R}$ and $\boldsymbol{\mu}$ can be obtained in closed-forms since problem~\eqref{prob:cov-est-dist-min-multiary-relaxed} is a quadratic programming problem with linear constraints, detailed in the next. 

        According to~\cite{ref:OLS}, each $N\times N$ Hermitian matrix $\boldsymbol{A}$ can be equivalently represented by an $N^2$-dimensional real vector $\boldsymbol{w}_a$, which is the coordinate of $\boldsymbol{A}$ in the Hermitian matrix space. 
        Let $\mathcal{M}:\mathbb{C}^{N\times N}\to\mathbb{R}^{N^2\times 1}$ be the bijective mapping from $\boldsymbol{A}$ to $\boldsymbol{w}_a$, such that $\boldsymbol{w}_a = \mathcal{M}(\boldsymbol{A})$ and $\boldsymbol{A} = \mathcal{M}^{-1}(\boldsymbol{w}_a)$. 
        It has been shown in~\cite{ref:OLS} that for Hermitian matrices $\boldsymbol{A}$ and $\boldsymbol{B}$, we have $\textup{tr}(\boldsymbol{A}\boldsymbol{B}) = \boldsymbol{w}_a^T\boldsymbol{w}_b$, where $\boldsymbol{w}_a = \mathcal{M}(\boldsymbol{A})$ and $\boldsymbol{w}_b = \mathcal{M}(\boldsymbol{B})$. 
        Denote $\boldsymbol{w} = \mathcal{M}(\boldsymbol{R})$, $\boldsymbol{\phi}_{l} = \mathcal{M}(\boldsymbol{x}_{l}\boldsymbol{x}_{l}^{H})$, $\forall l$, $\boldsymbol{\psi}_t = \mathcal{M}(\boldsymbol{V}_t)$, $\forall t$, and matrices $\boldsymbol{\Phi} = [\boldsymbol{\phi}_{1}, \ldots, \boldsymbol{\phi}_{D}]\in\mathbb{R}^{N^{2}\times D}$ and $\boldsymbol{C} = {P}_{0}[\boldsymbol{\psi}_{1}, \ldots, \boldsymbol{\psi}_{T_{p}}]\in\mathbb{R}^{N^{2}\times{T_{p}}}$. 
        Then, we have
        \begin{equation}
            \varphi_{k}\left(
                \boldsymbol{R}, \boldsymbol{\mu}, \boldsymbol{X}
            \right) = \|
                \boldsymbol{w} - \boldsymbol{\Phi}\boldsymbol{\mu}
            \|_{2}^{2} + \rho\|
                \boldsymbol{C}^{T}\boldsymbol{w} - \boldsymbol{\beta}_{k}
            \|_{2}^{2}, 
        \end{equation}
        where vector $\boldsymbol{\beta}_{k} = [\beta_{k, 1}, \ldots, \beta_{k, T_{p}}]^{T}\in\mathbb{R}^{T_{p}\times 1}$ is defined as $\beta_{k, t} = q_{k, t} - M\sigma^{2}$, $\forall k, t$. 
        Hence, problem~\eqref{prob:cov-est-dist-min-multiary-relaxed} given $\boldsymbol{X}$ can be equivalently rewritten as 
        \begin{equation}\label{prob:cov-est-dist-min-multiary-decomp}
            \mathop{\min_{\boldsymbol{w}, \boldsymbol{\mu}}}\ \|
                \boldsymbol{w} - \boldsymbol{\Phi}\boldsymbol{\mu}
            \|_{2}^{2} + \rho\|
                \boldsymbol{C}^{T}\boldsymbol{w} - \boldsymbol{\beta}_{k}
            \|_{2}^{2}, 
        \end{equation}
        where constraint $\boldsymbol{R}^{H} = \boldsymbol{R}$ is guaranteed by the mapping $\mathcal{M}$ and thus is omitted. 
        By taking the gradient of the objective function with respect to $\boldsymbol{w}$ and $\boldsymbol{\mu}$, the optimal solutions for problem~\eqref{prob:cov-est-dist-min-multiary-decomp}, denoted by $\boldsymbol{w}^{\star}$ and $\boldsymbol{\mu}^{\star}$, are obtained as
        \begin{subequations}\label{def:dist-min-multiary-decomp-optimal-solutions}
            \begin{align}
                \boldsymbol{w}^{\star} & = \boldsymbol{\Upsilon}\boldsymbol{\Phi}\boldsymbol{\mu}^{\star} + \boldsymbol{\chi}_{k}, ~~ \boldsymbol{\mu}^{\star} = \big(
                    \boldsymbol{Q}^{T}\boldsymbol{Q}
                \big)^{\dagger}\boldsymbol{Q}^{T}\boldsymbol{\chi}_{k}, \\
                \boldsymbol{\chi}_{k} & = \rho\boldsymbol{\Upsilon}\boldsymbol{C}\boldsymbol{\beta}_{k}, ~~ \boldsymbol{Q} = \big(
                    \boldsymbol{I}_{N^{2}} - \boldsymbol{\Upsilon}
                \big)\boldsymbol{\Phi}, 
            \end{align}
        \end{subequations}
        where $\boldsymbol{\Upsilon} = (\boldsymbol{I}_{N^{2}} + \rho\boldsymbol{C}\boldsymbol{C}^{T})^{-1}$. 
        The optimal solution of $\boldsymbol{R}$ for problem~\eqref{prob:cov-est-dist-min-multiary-relaxed} given $\boldsymbol{X}$ can be obtained as $\boldsymbol{R}^{\star} = \mathcal{M}^{-1}(\boldsymbol{w}^{\star})$. 
        
        \vspace{-8pt}
        \begin{algorithm}[h]
            \begin{minipage}{0.95\linewidth}
            \centering
            \caption{W-ALRA to solve $\mathcal{P}_{b}^{k}(D)$ for $b \ge 2$. }
            \begin{algorithmic}[1]\label{alg:W-ALRA-multiary-robust}
                \REQUIRE $\boldsymbol{v}_t$, $q_{k, t}$, $\forall t$; $\sigma^2$, $\rho$, $M$, and the iteration number $I$. 
                \STATE Solve $\boldsymbol{R}^{(0)}$ and $\boldsymbol{\mu}^{(0)}$ from problem~\eqref{prob:cov-est-dist-min-multiary-relaxed} given $\boldsymbol{X} = \boldsymbol{X}^{(0)} \triangleq \boldsymbol{0}_{N^{2}\times{D}}$ following equations~\eqref{def:dist-min-multiary-decomp-optimal-solutions}; index $i\gets 1$. 
                \WHILE{$i\le I$}
                    \STATE Let $\boldsymbol{x}_{l}^{(i)}$ be the normlized eigenvector of $\boldsymbol{R}^{(i - 1)}$ corresponding to the $l$-th largest eigenvalues, $1\le l\le D$, and $\boldsymbol{X}^{(i)} \gets [\boldsymbol{x}_{1}^{(i)}, \ldots, \boldsymbol{x}_{D}^{(i)}]$. 
                    \STATE Given $\boldsymbol{X} = \boldsymbol{X}^{(i)}$, solve $\boldsymbol{R}^{(i)}$ and $\boldsymbol{\mu}^{(i)}$ optimally from problem~\eqref{prob:cov-est-dist-min-multiary-relaxed} following equations~\eqref{def:dist-min-multiary-decomp-optimal-solutions}. 
                    \STATE $i\gets i + 1$. 
                \ENDWHILE
                \RETURN The estimated matrix $\hat{\boldsymbol{R}}_{c, k}\gets\boldsymbol{R}^{(i - 1)}$. 
            \end{algorithmic}
            \end{minipage}
        \end{algorithm}
        \vspace{-8pt}
        
        The proposed W-ALRA algorithm for $b\ge 2$ is summarized in Algorithm~\ref{alg:W-ALRA-multiary-robust}, where $\boldsymbol{R}^{\star}$ and $\boldsymbol{\mu}^{\star}$ solved from problem~\eqref{prob:cov-est-dist-min-multiary-relaxed} given $\boldsymbol{X} = \boldsymbol{0}_{N^{2}\times{D}}$ is employed for initialization. 
        To show the convergence of the algorithm, let $\lambda_{l}^{(i)}$ as the $l$-th largest eigenvalue of $\boldsymbol{R}^{(i - 1)}$, $\forall l$, and $\boldsymbol{\lambda}^{(i)} = [\lambda_{1}^{(i)}, \ldots, \lambda_{D}^{(i)}]^{T}\in\mathbb{R}^{D\times 1}$. 
        Then, the objective function of problem~\eqref{prob:cov-est-dist-min-multiary-relaxed} can be verified to be non-increasing as follows,
        \begin{subequations}\label{eqn:dist-min-multiary-convergence}
            \begin{align}
                \varphi_{k}(\boldsymbol{R}^{(i)}, \boldsymbol{\mu}^{(i)}, \boldsymbol{X}^{(i)}) & \le \varphi_{k}(\boldsymbol{R}^{(i - 1)}, \boldsymbol{\lambda}^{(i)}, \boldsymbol{X}^{(i)}) \label{subeq:dist-min-multiary-convergence-eigenvalue} \\
                & \le \varphi_{k}(\boldsymbol{R}^{(i - 1)}, \boldsymbol{\mu}^{(i - 1)}, \boldsymbol{X}^{(i - 1)}), \label{subeq:dist-min-multiary-convergence-eigenvector}
            \end{align}
        \end{subequations}
        where~\eqref{subeq:dist-min-multiary-convergence-eigenvalue} results from the optimality of $\boldsymbol{R}^{(i)}$ and $\boldsymbol{\mu}^{(i)}$ given $\boldsymbol{X}^{(i)}$ while~\eqref{subeq:dist-min-multiary-convergence-eigenvector} utilizes the fact that $\boldsymbol{\lambda}^{(i)}$ and $\boldsymbol{X}^{(i)}$ optimally solve problem~\eqref{prob:cov-est-dist-min-multiary-relaxed} given $\boldsymbol{R}^{(i - 1)}$. 
        Moreover, since $\boldsymbol{\Upsilon}$ can be calculated offline given $\boldsymbol{C}$ and $\rho$, the online computational complexity of Algorithm~\ref{alg:W-ALRA-multiary-robust} is dominated by that of the multiplication by $\boldsymbol{\Upsilon}$ to vectors, which is given by $\mathcal{O}(N^{2}T_{p}I)$ by rewriting $\boldsymbol{\Upsilon} = \boldsymbol{I}_{N^{2}} - \boldsymbol{C}(\rho^{-1}\boldsymbol{I}_{T_{p}} + \boldsymbol{C}^{T}\boldsymbol{C})^{-1}\boldsymbol{C}^{T}$.

    \vspace{-6pt}
    \subsection{Case of $b = 1$ given $D$}
    \label{subsec:case-binary}
        For $b = 1$, only the real part of $\boldsymbol{R}_{c, k}$, i.e., $\boldsymbol{R}_{r, k}$, is estimated, $\forall k$. 
        Following the same penalty and relaxation as the case of $b\ge 2$, problem $\mathcal{P}_{1}^{k}(D)$ given in~\eqref{prob:channel-autocorr-est-binary} can be approximated as
        \begin{subequations}\label{prob:cov-est-dist-min-binary-relaxed}
            \allowdisplaybreaks
            \begin{align}
                & \mathop{\min_{\boldsymbol{R}_{r}, \boldsymbol{\mu}_{r}, \boldsymbol{X}_{r}}}\ \varphi_{k}\left(
                    \boldsymbol{R}_{r}, \boldsymbol{\mu}_{r}, \boldsymbol{X}_{r}
                \right) \tag{\ref{prob:cov-est-dist-min-binary-relaxed}} \\
                & ~ \mathrm{s.t.} \ \boldsymbol{R}_{r}^{T} = \boldsymbol{R}_{r}, ~\boldsymbol{X}_{r}^{T}\boldsymbol{X}_{r} = \boldsymbol{I}_{2D}, \label{prob:cov-est-dist-min-binary-relaxed-hermitian}
            \end{align}
        \end{subequations}
        where we have $\boldsymbol{\mu}_{r} = [\mu_{r, 1}, \ldots, \mu_{r, 2D}]\in\mathbb{R}^{2D\times 1}$ and $\boldsymbol{X}_{r} = [\boldsymbol{x}_{r,1}, \ldots, \boldsymbol{x}_{r,2D}]\in\mathbb{R}^{N^{2}\times{2D}}$ with $\boldsymbol{x}_{r, l}\in\mathbb{R}^{N^{2}\times 1}$, $1\le l\le 2D$. 
        Similar to the case of $b\ge 2$, AO is employed to solve problem~\eqref{prob:cov-est-dist-min-binary-relaxed} sub-optimally. 
        Given $\boldsymbol{R}_{r}$, the optimal $\mu_{r, l}$ and $\boldsymbol{x}_{r, l}$ can be obtained as the $l$-th largest eigenvalue of $\boldsymbol{R}_{r}$ and the corresponding normalized eigenvector, respectively, $\forall l$. 
        Given $\boldsymbol{X}_{r}$, define $\boldsymbol{w}_{r} = \mathcal{M}(\boldsymbol{R}_{r})$, $\boldsymbol{\phi}_{r, l} = \mathcal{M}(\boldsymbol{x}_{r, l}\boldsymbol{x}_{r, l}^{T})$, $\forall l$, $\boldsymbol{\Phi}_{r} = [\boldsymbol{\phi}_{r, 1}, \ldots, \boldsymbol{\phi}_{r, 2D}]\in\mathbb{R}^{N^{2}\times 2D}$, and matrix $\boldsymbol{C}$ similarly as for $b\ge 2$. 
        The optimal solution for problem~\eqref{prob:cov-est-dist-min-binary-relaxed} given $\boldsymbol{X}_{r}$ can be obtained as $\boldsymbol{R}_r^{\star} = \mathcal{M}^{-1}(\boldsymbol{w}_r^{\star})$ and $\boldsymbol{\mu}_{r}^{*}$, where $\boldsymbol{w}_r^{\star}$ and $\boldsymbol{\mu}_{r}^{*}$ are obtained by solving 
        \begin{equation}\label{prob:dist-min-approx-robust-binary-decomp}
            \mathop{\min_{{\boldsymbol{w}_r}, \boldsymbol{\mu}_{r}}}\ \|{\boldsymbol{w}_r} - \boldsymbol{\Phi}_{r}\boldsymbol{\mu}_{r}\|_2^2 + \rho\|{\boldsymbol{C}}^T{\boldsymbol{w}_r} - \boldsymbol{\beta}_{k}\|_2^2. 
        \end{equation}
        As such, ${\boldsymbol{w}}_r^{\star}$ and $\boldsymbol{\mu}^{\star}$ can be obtained as 
        \begin{equation}\label{def:dist-min-binary-decomp-optimal-solutions}
            {\boldsymbol{w}}_r^{\star} = \boldsymbol{\Upsilon}\boldsymbol{\Phi}_{r}\boldsymbol{\mu}_{r}^{\star} + \boldsymbol{\chi}_{k}, 
            ~~ \boldsymbol{\mu}_{r}^{\star} = \rho\big(
                \boldsymbol{Q}_{r}^{T}\boldsymbol{Q}_{r}
            \big)^{\dagger}\boldsymbol{Q}_{r}^{T}\boldsymbol{\chi}_{k}, 
        \end{equation}
        where $\boldsymbol{Q}_{r} = (\boldsymbol{I}_{N^2} - \boldsymbol{\Upsilon})\boldsymbol{\Phi}_{r}$, while $\boldsymbol{\chi}_{k}$ and $\boldsymbol{\Upsilon}$ are the same as in the $b\ge 2$ case. 
        Thus, the W-ALRA algorithm for $b = 1$ can be implemented similarly to Algorithm~\ref{alg:W-ALRA-multiary-robust}, where $\boldsymbol{X}_{r}$ and $\boldsymbol{R}_{r}$, $\boldsymbol{\mu}_{r}$ are optimized alternately. 
        The initialization is set as $\boldsymbol{R}_r^{\star}$ and $\boldsymbol{\mu}_{r}^{\star}$ with $\boldsymbol{X}_{r} = \boldsymbol{0}_{N^2\times 2D}$. 

    \vspace{-6pt}
    \subsection{Progressive Refinement of $D$}
    \label{subsec:progressive-refine}
        Since the W-ALRA algorithm estimates the channel autocorrelation matrix by leveraging its low-rank property, a suitable value for $D$ is crucial to obtain better estimations. 
        Despite that $D_{k}$ is usually difficult to obtain in practice, $D$ can be progressively refined to approximate $D_{k}$ more closely. 
        Specifically, we propose to solve problem $\mathcal{P}_{b}^{k}(D)$ by applying W-ALRA progressively with increasing value of $D$, i.e., $D = 1, 2, \ldots, M$. 
        The channel autocorrelation matrix solved in the previous iteration can be employed as the initialization for the next one, such that the estimation is gradually refined along with $D$. 
        Note that the estimated channel autocorrelation matrix should be almost the same for problem $\mathcal{P}_{b}^{k}(D)$ with $D\ge D_{k}$. 
        Thus, the refinement terminates once the relative error between the estimated channel autocorrelation matrices of two consecutive iterations is smaller than a given threshold $\varepsilon$.

\vspace{-4pt}
\section{Performance Evaluation}\label{sec:performance-evaluation}
    \vspace{-2pt}
    \subsection{Simulation Setup}\label{subsec:perf-setup}
\begin{figure}[t!]
            \centering
            {
                \begin{subfigure}[t]{0.235\textwidth}
                    \centering
                    \includegraphics[scale = 0.28]{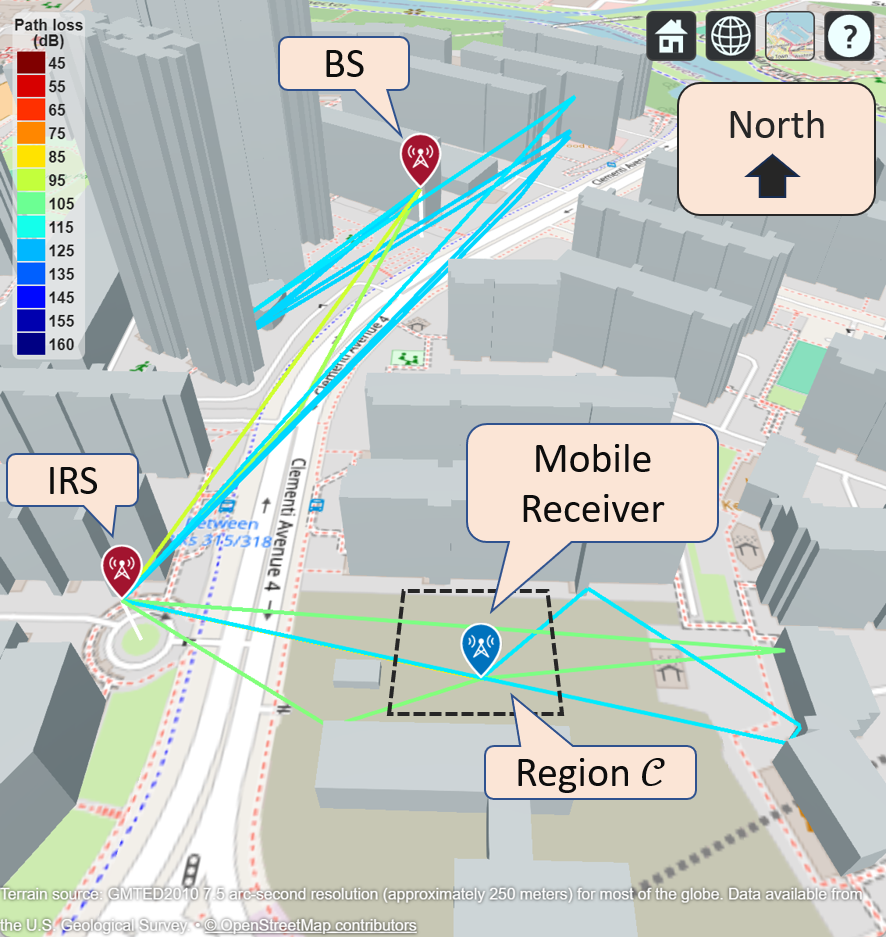}
                    \caption{Simulation setup. }
                    \label{subfig:setup-view}
                \end{subfigure}
                \begin{subfigure}[t]{0.235\textwidth}
                    \centering
                    \includegraphics[scale = 0.28]{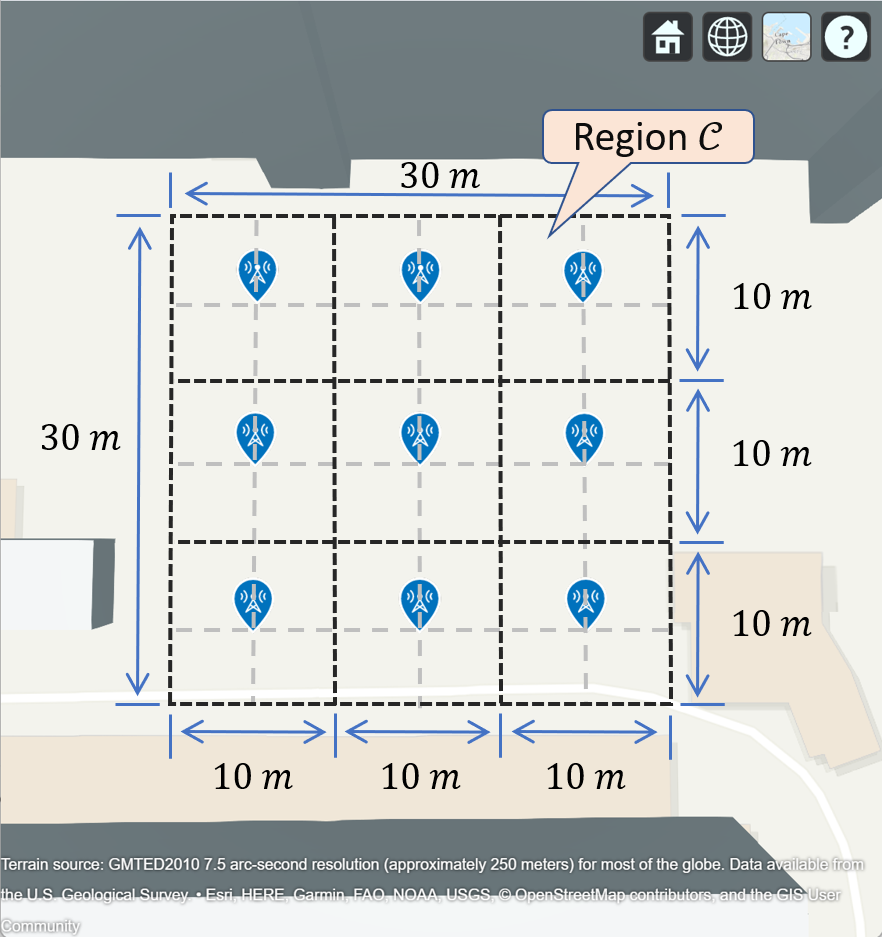}
                    \caption{Sampled locations. }
                    \label{subfig:sampled-grids}
                \end{subfigure}
            }
            \vspace{-2pt}
            \caption{Simulation setup and an example of sampled locations. }
            \label{fig:setup}
            \vspace{-15pt}
        \end{figure}

        To validate the proposed channel autocorrelation matrix estimation and coverage enhancement approach, we apply it to an IRS-aided system over a region of interest $\mathcal{C}$ according to the urban map data at Clementi, Singapore~\cite{ref:openstreetmap}. 
        As shown in Fig.~\ref{subfig:setup-view}, the BS and the IRS are represented by red markers, where the direct link between the BS and the square region $\mathcal{C}$ of size $30\times 30$ meters is blocked by buildings. 
        The BS is $25$ meters high, facing to the South, while a $15$-meters-high IRS equipped with $N = 64$ reflecting elements is placed $70.78$ meters to the West and $227.34$ meters to the South of the BS, forming an $8\times 8$ passive array and facing to the East. 
        Besides, the center of $\mathcal{C}$ is on the ground, $95.45$ meters to the East of the IRS. 
        For simplicity, $K = K_{0}^{2}$ locations are sampled as regular $K_{0}\times K_{0}$ grids within $\mathcal{C}$, where $K_{0}$ denotes the number of sampled locations along one side of $\mathcal{C}$. 
        In Fig.~\ref{subfig:sampled-grids}, an example is shown for $K = 3\times 3$. 
        
        The carrier frequency of the OFDM signal is $f_{c} = 1.4$ GHz, with $M = 128$ subcarriers employed over a total bandwidth of $20$ MHz. 
        Besides, we set transmit power ${P}_{0} = 30$ dBm, noise power $\sigma^{2} = -120$ dBm for each subcarrier, $J = 8$, $I = 20$, $\rho = 10$, and $\varepsilon = 0.005$. 
        Unless otherwise stated, $K = 3\times 3 = 9$ locations are sampled within $\mathcal{C}$ as in Fig.~\ref{subfig:sampled-grids}. 
        The wideband channels for the BS-IRS and IRS-receiver links are generated via ray-tracing for all sampled locations, which are obtained by taking into account the delay of each multipath component. 
        According to ray-tracing, the maximum delay spread within $\mathcal{C}$ for the BS-IRS-receiver link is $\tau = 5.6$ $\mathrm{\mu}$s, which yields $L = 112$. 
        The number of paths for the BS-IRS-receiver link varies between $4$ and $20$ within the region. 
        The IRS reflection vectors for training are randomly generated with each element independently and uniformly distributed in $\Phi_{b}$. 
        
    \vspace{-8pt}
    \subsection{Numerical Results}\label{subsec:perf-convergence-and-est-error}
\begin{figure}[t!]
            \centering
            {
                \begin{subfigure}[t]{0.19\textwidth}
                    \centering
                    \includegraphics[scale = 0.42]{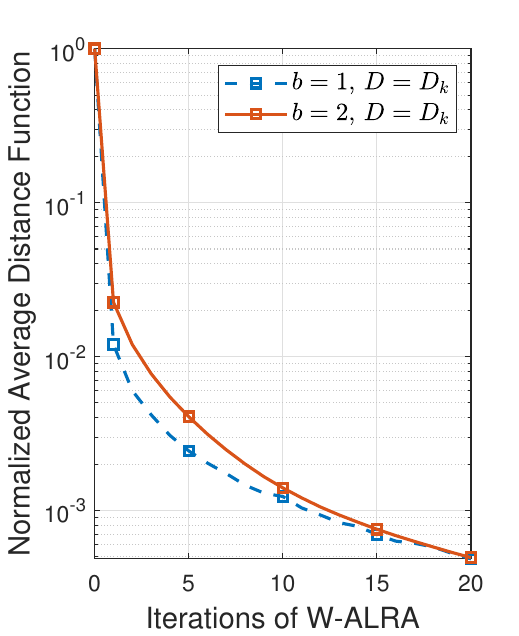}
                    \caption{Convergence of W-ALRA with $T_{p} = 96$. }
                    \label{subfig:alg-conv}
                \end{subfigure}
                \hspace{-5pt}
                \begin{subfigure}[t]{0.28\textwidth}
                    \centering
                    \includegraphics[scale = 0.42]{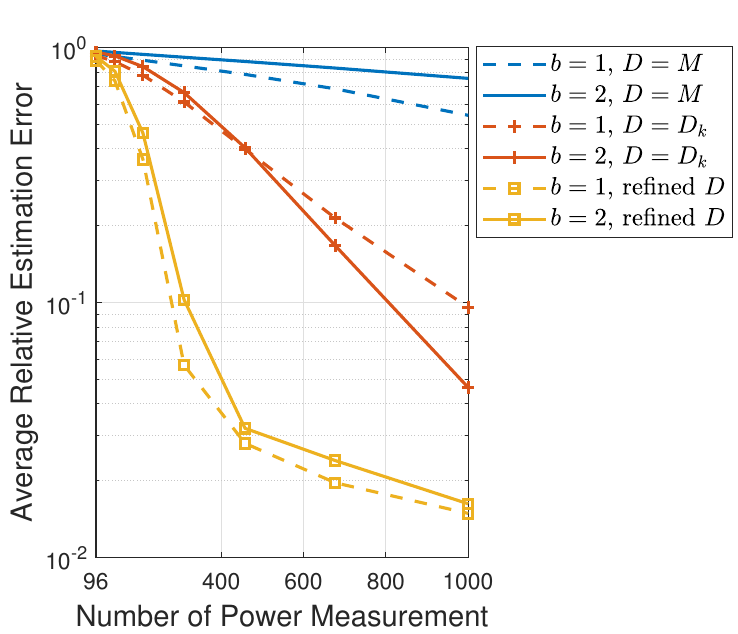}
                    \caption{Estimation error for W-ALRA with various choices for $D$. }
                    \label{subfig:est-err}
                \end{subfigure}
                
            }
            \vspace{-2pt}
            \caption{Convergence and estimation error of W-ALRA. }
            \label{fig:conv-and-est-err}
            \vspace{-15pt}
        \end{figure}

        To verify the convergence of proposed W-ALRA algorithm, the penalized distance function values during the iterations, i.e., $\varphi_{k}(\boldsymbol{R}^{(i)}, \boldsymbol{\mu}^{(i)}, \boldsymbol{X}^{(i)})$ for $b = 2$ and $\varphi_{k}(\boldsymbol{R}_{r}^{(i)}, \boldsymbol{\mu}_{r}^{(i)}, \boldsymbol{X}_{r}^{(i)})$ for $b = 1$, are averaged over all $K$ sampled locations and shown in Fig.~\ref{subfig:alg-conv} after normalization with $T_{p} = 96$, where W-ALRA is applied with $D = D_{k}$, $\forall k$. 
        It can be observed that the objective function values keep decreasing, indicating that the algorithm gradually approaches a rank-$D$ ($2D$ for $b = 1$) solution. 
        Besides, define relative estimation error $\mathcal{E}_{k}^{b} = \|\hat{\boldsymbol{R}}_{k} - \boldsymbol{R}_{k}\|_{F}/\|\boldsymbol{R}_{k}\|_{F}$ for $b\ge 2$ and $\mathcal{E}_{k}^{1} = \|\hat{\boldsymbol{R}}_{r, k} - \boldsymbol{R}_{r, k}\|_{F}/\|\boldsymbol{R}_{r, k}\|_{F}$ for $b = 1$, $\forall k$. 
        In Fig.~\ref{subfig:est-err}, the averaged relative estimation error $\bar{\mathcal{E}}_{b} = \frac{1}{K}\sum_{k = 1}^{K}{\mathcal{E}_{k}^{b}}$ is shown for W-ALRA with $D = M$, $D = D_{k}$, $\forall k$, and progressive refinement of $D$, respectively. 
        With $D = M$, the error is quite large because the rank of the channel autocorrelation matrix is significantly overestimated. 
        In contrast, the estimation errors for W-ALRA with known $D_{k}$'s and progressive refinement for $D$ are much lower. 
        In particular, it is interesting to see that the W-ALRA with progressive refinement of $D$ achieves lower estimation error than that with $D = D_{k}$, $\forall k$, which is due to the fact that better initializations are adopted for W-ALRA over the refinement. 

\begin{figure}[t]
        \begin{center}
            \includegraphics[scale = 0.41]{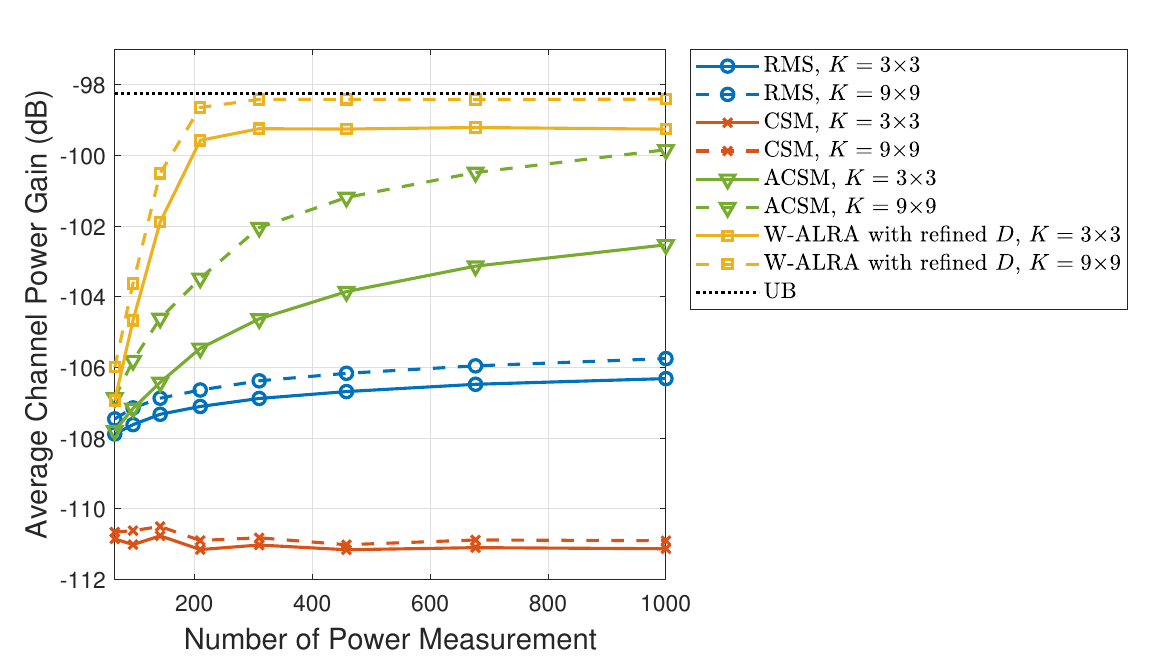}
            \vspace{-1.5pt}
            \caption{Average channel power gain versus $T_{p}$ for the proposed and benchmark schemes with $b = 2$. }
            \label{fig:channel-gain}
        \end{center}
        \vspace{-20pt}
    \end{figure}

        After estimating channel autocorrelation matrices and optimizing the IRS reflection vector as $\hat{\boldsymbol{v}}$ based on them, the average channel power gain $\bar{\gamma} = \hat{\boldsymbol{v}}^{H}\mathbb{E}[\boldsymbol{R}_{c}]\hat{\boldsymbol{v}}$ over the region is computed, where $\mathbb{E}[\boldsymbol{R}_{c}]$ is approximated by averaging $500$ channel autocorrelation matrix realizations within $\mathcal{C}$. 
        In Fig.~\ref{fig:channel-gain}, simulation results for $\bar{\gamma}$ are shown for the case of $b = 2$, where the performance obtained by the proposed W-ALRA algorithm is labeled as `W-ALRA'. 
        We show the performance results for both $K = 3\times 3$ and $K = 9\times 9$ sampled locations.  
        For comparison, four benchmark schemes are considered: 
        i) \textbf{UB} (upper bound): The upper bound on the average channel power gain is obtained by the IRS reflection vector optimized assuming nearly perfect knowledge of $\mathbb{E}[\boldsymbol{R}_{c}]$, achieved by employing the empirical average channel autocorrelation matrix over $500$ (instead of $K$) sampled locations; 
        ii) \textbf{RMS} (random-max sampling): Random IRS reflection vectors are applied with $v_{n}$ uniformly distributed in $\Phi_b$, $\forall n$, and the one achieving the largest average channel gain over sampled locations is used; 
        iii) \textbf{CSM} (conditional sample mean): This is the method proposed in~\cite{ref:CSM}, where each phase shift is selected to maximize the empirical expectation of average channel gain conditioned on it. 
        iv) \textbf{ACSM} (adaptive CSM): The extension of CSM proposed in~\cite{ref:acsm} for cases with weak/no direct links, which refines the IRS reflections in two stages and applies CSM alternately with part of IRS phase shifts fixed. 

        As shown in Fig.~\ref{fig:channel-gain}, the average channel power gain obtained by W-ALRA with progressive refinement increases rapidly as $T_{p}$ grows, significantly outperforming all benchmark schemes. 
        Note that the gap between W-ALRA and UB for large $T_{p}$ is due to that $\hat{v}$ is optimized based on the average autocorrelation matrix over $K$ sampled locations, which cannot exactly represent the whole region $\mathcal{C}$. 
        Moreover, the average channel power gain achieved by each scheme with $K = 9\times 9$ is always higher than that with $3\times 3$ because the former can collect more power measurements and thus acquire more accurate channel autocorrelation knowledge in the region. 
        Nevertheless, even with $K = 3\times 3$, W-ALRA still closely approaches the upper bound for $T_{p} \ge 300$, which is not achievable even for $T_{p} = 1000$ for other benchmark schemes with $K = 9\times 9$. 
        These results validate the effectiveness of our proposed approach for IRS reflection design based on channel autocorrelation matrix estimation in wideband/OFDM communication systems.

\vspace{-4pt}
\section{Conclusion}\label{Conclusion}
    This letter investigated the channel autocorrelation matrix estimation and coverage enhancement for an IRS-aided OFDM system based on received signal power measurement, withouit the need for pilot-assisted instantaneous channel estimation. 
    By dividing the region of interest into orthognal blocks and estimating the channel autocorrelation matrix at the center of each selected block, the average channel gain over the region can be improved by designing the IRS reflection vector based on the estimated channel autocorrelation matrices. 
    To verify the effectiveness of the proposed W-ALRA algorithm, simulations were conducted under practical ray-tracing channels. 
    The results demonstrated the high estimation accuracy of the W-ALRA algorithm and its resulting coverage improvement within the region compared to benchmark schemes, which offers a practically appealing approach for quasi-static IRS reflection design in dynamic and multi-path environments with frequency-selective fading channels.

\bibliographystyle{IEEEtran} 
\bibliography{IEEEabrv, reference}

\end{document}